\documentclass[twoside]{article}

\usepackage{amsmath}
\usepackage{amssymb}
\usepackage{epsf}

\catcode`\@=11
\long\def\@makefntext#1{
\protect\noindent \hbox to 3.2pt {\hskip-.9pt
$^{{\eightrm\@thefnmark}}$\hfil}#1\hfill}       

\def\@makefnmark{\hbox to 0pt{$^{\@thefnmark}$\hss}}    

\def\ps@myheadings{\let\@mkboth\@gobbletwo
\def\@oddhead{\hbox{}
\rightmark\hfil\eightrm\thepage}
\def\@oddfoot{}\def\@evenhead{\eightrm\thepage\hfil
\leftmark\hbox{}}\def\@evenfoot{}
\def\sectionmark##1{}\def\subsectionmark##1{}}

\oddsidemargin=\evensidemargin
\newcounter{sectionc}\newcounter{subsectionc}\newcounter{subsubsectionc}
\renewcommand{\section}[1] {\vspace{12pt}\addtocounter{sectionc}{1}
\setcounter{subsectionc}{0}\setcounter{subsubsectionc}{0}\noindent
    {\tenbf\thesectionc. #1}\par\vspace{5pt}}
\renewcommand{\subsection}[1] {\vspace{12pt}\addtocounter{subsectionc}{1}
    \setcounter{subsubsectionc}{0}\noindent
        {\tenbf\thesectionc.\thesubsectionc. {\kern1pt\bfit #1}}
    \par\vspace{5pt}}
\renewcommand{\subsubsection}[1] {\vspace{12pt}\addtocounter{subsubsectionc}{1}
    \noindent{\tenrm\thesectionc.\thesubsectionc.\thesubsubsectionc.
    {\kern1pt \tenit #1}}\par\vspace{5pt}}

\newcounter{appendixc}
\newcounter{subappendixc}[appendixc]
\newcounter{subsubappendixc}[subappendixc]
\renewcommand{\thesubappendixc}{\Alph{appendixc}.\arabic{subappendixc}}
\renewcommand{\thesubsubappendixc}
    {\Alph{appendixc}.\arabic{subappendixc}.\arabic{subsubappendixc}}

\renewcommand{\appendix}[1] {\vspace{12pt}
        \refstepcounter{appendixc}
        \setcounter{figure}{0}
        \setcounter{table}{0}
        \setcounter{lemma}{0}
        \setcounter{theorem}{0}
        \setcounter{corollary}{0}
        \setcounter{definition}{0}
        \setcounter{equation}{0}
        \renewcommand{\thefigure}{\Alph{appendixc}.\arabic{figure}}
        \renewcommand{\thetable}{\Alph{appendixc}.\arabic{table}}
        \renewcommand{\theappendixc}{\Alph{appendixc}}
        \renewcommand{\thelemma}{\Alph{appendixc}.\arabic{lemma}}
        \renewcommand{\thetheorem}{\Alph{appendixc}.\arabic{theorem}}
        \renewcommand{\thedefinition}{\Alph{appendixc}.\arabic{definition}}
        \renewcommand{\thecorollary}{\Alph{appendixc}.\arabic{corollary}}
        \renewcommand{\theequation}{\Alph{appendixc}.\arabic{equation}}
        \noindent{\tenbf Appendix \theappendixc #1}\par\vspace{5pt}}
\newcommand{\subappendix}[1] {\vspace{12pt}
        \refstepcounter{subappendixc}
        \noindent{\bf Appendix \thesubappendixc. {\kern1pt \bfit #1}}
    \par\vspace{5pt}}
\newcommand{\subsubappendix}[1] {\vspace{12pt}
        \refstepcounter{subsubappendixc}
        \noindent{\rm Appendix \thesubsubappendixc. {\kern1pt \tenit #1}}
    \par\vspace{5pt}}

\newcommand{\textlineskip}{\baselineskip=13pt}
\newcommand{\smalllineskip}{\baselineskip=10pt}

\def\eightcirc{
\begin{picture}(0,0)
\put(4.4,1.8){\circle{6.5}}
\end{picture}}
\def\eightcopyright{\eightcirc\kern2.7pt\hbox{\eightrm c}}

\newcommand{\copyrightheading}[1]
    {\vspace*{-2.5cm}\smalllineskip{\flushleft
    {\footnotesize International Journal of Uncertainty, Fuzziness
     and Knowledge-Based Systems\\
     #1}\\
    {\footnotesize $\eightcopyright$\, World Scientific Publishing
     Company}\\
     }}

\newcommand{\publisher}[2]{{\begin{center}\footnotesize\smalllineskip
    Received #1\\
    Revised #2\end{center}
    }}

\def\abstracts#1#2#3{{
    \centering{\begin{minipage}{4.5in}\baselineskip=10pt\footnotesize
    \parindent=0pt #1\par
    \parindent=15pt #2\par
    \parindent=15pt #3
    \end{minipage}}\par}}

\def\keywords#1{{
    \centering{\begin{minipage}{4.5in}\baselineskip=10pt\footnotesize
    {\footnotesize\it Keywords}\/: #1
    \end{minipage}}\par}}

\newcounter{itemlistc}
\newcounter{romanlistc}
\newcounter{alphlistc}
\newcounter{arabiclistc}

\newcommand{\fcaption}[1]{
        \refstepcounter{figure}
        \setbox\@tempboxa = \hbox{\footnotesize Fig.~\thefigure. #1}
        \ifdim \wd\@tempboxa > 5in
           {\begin{center}
        \parbox{5in}{\footnotesize\smalllineskip Fig.~\thefigure. #1}
            \end{center}}
        \else
             {\begin{center}
             {\footnotesize Fig.~\thefigure. #1}
              \end{center}}
        \fi}

\newcommand{\tcaption}[1]{
        \refstepcounter{table}
        \setbox\@tempboxa = \hbox{\footnotesize Table~\thetable. #1}
        \ifdim \wd\@tempboxa > 5in
           {\begin{center}
        \parbox{5in}{\footnotesize\smalllineskip Table~\thetable. #1}
            \end{center}}
        \else
             {\begin{center}
             {\footnotesize Table~\thetable. #1}
              \end{center}}
        \fi}

\def\@citex[#1]#2{\if@filesw\immediate\write\@auxout
    {\string\citation{#2}}\fi
\def\@citea{}\@cite{\@for\@citeb:=#2\do
    {\@citea\def\@citea{,}\@ifundefined
    {b@\@citeb}{{\bf ?}\@warning
    {Citation `\@citeb' on page \thepage \space undefined}}
    {\csname b@\@citeb\endcsname}}}{#1}}

\newif\if@cghi
\def\cite{\@cghitrue\@ifnextchar [{\@tempswatrue
    \@citex}{\@tempswafalse\@citex[]}}
\def\citelow{\@cghifalse\@ifnextchar [{\@tempswatrue
    \@citex}{\@tempswafalse\@citex[]}}
\def\@cite#1#2{{$\null^{#1}$\if@tempswa\typeout
    {IJCGA warning: optional citation argument
    ignored: `#2'} \fi}}

\def\pmb#1{\setbox0=\hbox{#1}
    \kern-.025em\copy0\kern-\wd0
    \kern.05em\copy0\kern-\wd0
    \kern-.025em\raise.0433em\box0}

\def\fnt#1#2{\footnotetext{\kern-.3em           
    {$^{\mbox{\sixrm\it #1}}$}{#2}}}

\def\fpage#1{\begingroup
\voffset=.3in
\thispagestyle{empty}\begin{table}[b]\centerline{\footnotesize #1}
    \end{table}\endgroup}

\def\runninghead#1#2{\pagestyle{myheadings}
\markboth{{\protect\footnotesize\it{\quad #1}}\hfill}
{\hfill{\protect\footnotesize\it{#2\quad}}}}
\font\tenrm=cmr10
\font\tenit=cmti10
\font\tenbf=cmbx10
\font\bfit=cmbxti10 at 10pt
\font\ninerm=cmr9

\font\eightrm=cmr8

\newtheorem{theorem}{\indent Theorem}

\newtheorem{lemma}{Lemma}

\newtheorem{definition}{Definition}
\newtheorem{corollary}{Corollary}
\newtheorem{proposition}{\indent Proposition}
\newtheorem{remark}{\indent Remark}
\newtheorem{example}{\indent Example}

\def\cqd{\hfill \rule{2.25mm}{2.25mm}\vspace{10pt}}

\def\qed{\hbox{${\vcenter{\vbox{            
   \hrule height 0.4pt\hbox{\vrule width 0.4pt height 6pt
   \kern5pt\vrule width 0.4pt}\hrule height 0.4pt}}}$}}

\def\bsc{{\sc a\kern-6.4pt\sc a\kern-6.4pt\sc a}}   
\def\bflatex{\bf L\kern-.30em\raise.3ex\hbox{\bsc}\kern-.14em
T\kern-.1667em\lower.7ex\hbox{E}\kern-.125em X}

\begin{document}

\runninghead{$p$-symmetric fuzzy measures} {$p$-symmetric fuzzy measures}

\normalsize\textlineskip
\thispagestyle{empty}
\setcounter{page}{1}

\copyrightheading{Vol. 0, No. 0 (1993) 000---000}

\vskip 1.5 in

\fpage{1}

\centerline{\bf $p$-SYMMETRIC FUZZY MEASURES}         
\vspace*{0.37truein}
\centerline{\footnotesize PEDRO MIRANDA}
\centerline{\footnotesize\it Universidad de Oviedo}
\centerline{\footnotesize\it Calvo Sotelo, s/n. 33007 Oviedo, Spain}
\baselineskip=10pt
\centerline{\footnotesize\it pmm@pinon.ccu.uniovi.es} 
\vspace*{10pt}
\centerline{\footnotesize MICHEL GRABISCH}
\vspace*{0.015truein}
\centerline{\footnotesize\it Universit\'e Pierre et Marie Curie- LIP6}
\centerline{\footnotesize\it  8, rue du Capitaine Scott, 75015 Paris (France)}
\centerline{\footnotesize\it Michel.Grabisch@lip6.fr}
\baselineskip=10pt
\vspace*{10pt}
\centerline{\footnotesize PEDRO GIL}
\centerline{\footnotesize\it Universidad de Oviedo}
\centerline{\footnotesize\it Calvo Sotelo, s/n. 33007 Oviedo, Spain}
\centerline{\footnotesize\it pedro@pinon.ccu.uniovi.es}
\baselineskip=10pt 
\vspace*{0.225truein}
\publisher{(received date)}{(revised date)}


\vspace*{0.21truein} \abstracts{ In this paper we propose a generalization of the concept of symmetric fuzzy measure based in a decomposition of the universal set in what we have called subsets of indifference. Some properties of these measures are studied, as well as their Choquet integral. Finally, a degree of interaction between the subsets of indifference is defined.}{}{}

\vspace*{5pt}

\vspace*{1pt}\textlineskip
\section{Introduction}
\vspace*{-0.5pt}
\noindent

Fuzzy measures are a generalization of probability measures for
which additivity is removed and monotonicity is imposed instead.
These measures have become a powerful tool in Decision Theory (see e.g. \cite{grro00}, the work of Schmeidler \cite{sch86} and \cite{chco00}); moreover, the
Choquet Expected Utility model generalizes the Expected Utility
one, and this model offers a simple theoretical foundation for
explaining phenomena that cannot be accounted for in the framework of
Expected Utility Theory, as the well known Ellsberg's and Allais' paradoxes (see \cite{chco00} for a survey about this topic).

However, the richness of fuzzy measures has its counterpart in the
complexity. If we deal with a space of $n$ elements, a probability measure
only needs $n-1$ coefficients, while a fuzzy measure needs $2^n-2$. In an attempt to decrease the
exponential complexity of fuzzy measures in practical
applications, Grabisch has introduced in \cite{gra97c} the concept of 
$k$-order additive measures or $k$-additive measures for short; $k$-additive measures are defined from the M\"obius
transform, and they can be represented by a
limited set of coefficients, at most $\sum_{i=1}^k {n\choose i}$. A characterization of $k$-additive measures based in Choquet Expected Utility Model can be found in \cite{migr00}.

On the other hand, it is a well known fact that an OWA operator \cite{yag88} is a discrete Choquet integral with respect to a symmetric fuzzy measure. Hence, Choquet integral generalizes OWA operators and, as before, the
richness of Choquet integral is paid by the complexity. Our goal is to introduce a
concept, similar to $k$-additive measures, bridging the gap between
symmetric fuzzy measures and general fuzzy measures. We propose a
definition of $p$-symmetry based in what we will call {\it subsets of indifference}, and we study some of their properties. 

Of course, Choquet integral with respect to a $p$-symmetric fuzzy measure generalizes the concept of OWA. Another generalization of OWA operators can be found in \cite{capr01}, in which it is defined the so-called {\it double aggregation operators} as an aggregation of two other aggregation operators.

The paper is organized as follows: In Section 2, we recall some basic concepts. Next, in Section 3, we give the definition of $p$-symmetric measures and study some of their properties. In Section 4, we study the expressions of $p$-symmetric measures for other representations of fuzzy measures. In Section 5 we deal with the Choquet integral of $p$-symmetric measures; in this section we also define a degree of interaction and study its relationship with the decomposition of Choquet integral. We finish with the conclusions and open problems.

\section{Notations and basic concepts}
\noindent

In the sequel, we will consider a finite universal set of $n$
elements, denoted $X=\{ x_1,...,x_n\} $. Subsets of $X$ are denoted with
capital letters $A, B,$ and so on, and also by $A_1,..., A_p$. The set of all subsets of $X$ is denoted ${\cal P}(X).$ Finally, $\bigwedge $ (resp. $\bigvee $) denotes the min (resp. max) operation.

In order to be self-contained, let us now give some definitions:

\begin{definition}\cite{sug74}
A (discrete) {\bf fuzzy measure} on $X$ is a set function
$ \mu:{\cal P}(X)\mapsto [0,1] $ satisfying
\begin{enumerate}
\item[(i)] $ \mu(\emptyset)=0,\quad\mu(X)=1 $ (boundary conditions).
\item[(ii)] $ A\subset B$  implies $ \mu (A)\leq\mu(B)$
(monotonicity).
\end{enumerate}
\end{definition}

To any fuzzy measure, we can assign another one, called dual measure whose definition is the following:

\begin{definition}
Consider $(X,{\cal X})$ a measurable space and let $\mu $ be a fuzzy measure over $X$; we define the {\bf dual} or {\bf conjugate measure of $\mu $} as the fuzzy measure $\bar{\mu }$ given by $ \bar{\mu }(A)=1-\mu (A^c),$ where $A^c=X\backslash A$. 
\end{definition}

Other alternative representations of fuzzy measures are given by the M\"obius transform and the Shapley interaction.

\begin{definition}\cite{rot64}
Let $ \mu $ be a fuzzy measure on X. The {\bf M\"{o}bius
transform} of  $\mu $ is defined by 
\begin{equation}\label{defmob}
m(A):=\sum_{B\subset A}(-1)^{|
A\backslash B|}\mu (B), \forall A\subset X.
\end{equation}
\end{definition}

In the Theory of Cooperative Games, the M\"obius transform is interpreted as the importance of each subset by itself, without considering its parts. In this sense, this transformation is called {\it dividend} \cite{dehape97}.

When $m $ is given, it is possible to recover the original $\mu $ by the so-called {\it Zeta transform} (see \cite{chja89}):
\begin{equation}\label{mob2mea}
\mu (A)=\sum_{B\subset A} m(B).
\end{equation}

We can define $m$ for any set function, not limited to fuzzy measures. In order to $m$ being the M\"obius transform of a fuzzy measure we need to impose some monotonicity constraints. These constraints are given in the following proposition:

\begin{proposition}\cite{chja89}\label{monconmob}
A set of $2^n$ coefficients $m(A), A\subset X$ corresponds to the M\"obius representation of a fuzzy measure if and only if
\begin{enumerate}
\renewcommand{\labelenumi}{(\roman{enumi})}
\item
$\displaystyle{ m(\emptyset )=0,\, \sum_{A\subset X} m(A)=1,}$
\item
$\displaystyle{\sum_{x_i\in B\subset A} m(B)\geq 0}$, $\forall A\subset X$, $\forall x_i\in A$.
\end{enumerate} 
\end{proposition}

>From M\"obius transform, we can derive the definition of belief function, given by Dempster \cite{dem67} and Shafer \cite{sha76}:

\begin{definition}
A fuzzy measure $\mu $ is a {\bf belief function} if $m(A)\geq 0,\forall A\subset X.$
\end{definition}

Shapley interaction is another equivalent representation of fuzzy measures.

\begin{definition}\label{sha}\cite{gra96c}
Let $\mu $ be a fuzzy measure on $X$. The {\bf Shapley interaction index} of $A\subset X $, is defined by:
$$ I_{\mu }(A)=\sum_{B\subset X\backslash A} {(n-b-a)! b!\over (n-a+1)!} \sum_{C\subset A} (-1)^{a-c} \mu (B\cup C),$$
with $a=|A|,\, b=|B|,\, c=|C|$.
\end{definition}

Shapley interaction for singletons is just the Shapley value of a game \cite{sha53}, and it recovers the interaction index of Murofushi and Soneda \cite{muso93} for pairs.
 
$I$ and $m$ are related through the following formulas:
\begin{equation}\label{mob2sha}
I(A)= \sum_{B\subset X\backslash A} {1\over |B|+1} m(A\cup B),\, \forall A\subset X.
\end{equation}
\begin{equation}\label{sha2mob}
m(A)= \sum_{B\subset X\backslash A} B_{|B|} I(B\cup A), \, \forall A\subset X,
\end{equation}
 where $B_{k}$ denotes the Bernoulli numbers defined by recurrence through  $B_0 =1$ and $B_k = -\sum_{l=0}^{k-1} {B_l\over k-l+1} {k\choose l}.$

\begin{definition}\cite{cho53}\label{choint}
The {\bf Choquet integral}\footnote{If $X$ is continuous, the measurability is needed. Let $(X,{\cal X})$ be a measurable space. We say a mapping $f$ is a measurable function if $\{ x|\, f(x)\geq \alpha \} $ is in the $\sigma $-algebra ${\cal X}$ for any $\alpha \geq 0$.} of a measurable function $
f:X\mapsto \mathbb{R}^+ $ is defined by $$ {\cal C}_{\mu }(f):=\int_0^{\infty }\mu(\{ x\vert f(x)\geq \alpha \} )\mbox{d}
\alpha .$$ For simple functions the expression reduces to: $$
{\cal C}_{\mu }(f):=\sum_{i=1}^n (f(x_{(i)})-f(x_{(i-1)}))\mu (B_i),$$ where parenthesis mean a
permutation such that $ 0=f(x_{(0)})\leq f(x_{(1)})\leq ...\leq f(x_{(n)}) $
and $ B_i=\{ x_{(i)},...,x_{(n)}\}. $
Another equivalent expression for simple functions is $$ {\cal C}_{\mu }(f) :=\sum_{i=1}^n
f(x_{(i)})(\mu (B_i)-\mu (B_{i+1})) $$ with $B_{n+1}=\emptyset $.
\end{definition}

Choquet integral in terms of $m$ is given by:

\begin{theorem}\cite{chja89}
The Choquet integral ${\cal C}_{\mu }:[0,1]^n\mapsto \mathbb{R}^+ $ can be
written as
\begin{equation}\label{chomob}
{\cal C}_{\mu }(f)=\sum_{T\subset X}m(T)\left[ \bigwedge_{x_i\in T} f(x_{i})\right] ,\quad
f\in [0,1]^n.
\end{equation}
\end{theorem}

\begin{definition}\cite{yag88}
An {\bf ordered weighted averaging operator (OWA)} is an operator defined by $$ \mathrm{OWA}_w(f)=\sum_{i=1}^nw_if(x_{(i)}),$$ where $w$ is the weight vector, $w=(w_1, ...,w_n)\in [0,1]^n$ and
such that $\sum_{i=1}^nw_i=1$.
\end{definition}

\begin{definition}
A fuzzy measure is said to be {\bf symmetric} if it satisfies $$ |A|=|B|\Rightarrow \mu (A)=\mu (B),\, \forall A, B\subset X.$$
\end{definition}

It can be proved (see \cite{gra96b} and \cite{musu93}) that:

\begin{proposition}\label{owasym}
Let $\mu $ be a fuzzy measure on $X$. Then, the following statements are
equivalent:
\begin{enumerate}
\item
There exists $w\in [0,1]^n,\, \sum_{i=1}^n w_i=1,$ such that ${\cal C}_{\mu }=\mbox{{\it
OWA}}_w.$
\item
$\mu $ is a symmetric fuzzy measure.
\end{enumerate}
\end{proposition}

Choquet integral model can be regarded as the generalization of a linear model in the sense that $({\cal C})\int f\, d\mu +({\cal C})\int g\, d\mu =({\cal C})\int (f+g)\, d\mu $ for a pair of comonotone functions \cite{halipo52}. The expressive power of Choquet integral is much higher than that of a linear model. However, as it can be seen from Definition \ref{choint}, Choquet integral model is difficult to handle. These are the reasons for which it has been proposed a {\it hierarchical Choquet integral model}, that allows to compute Choquet integral from combinations of other Choquet integrals. The underlying idea here is to be able to decompose the integral into a sum of other integrals over smaller referential sets. 

\begin{definition}\cite{musufu97}
Let $(X,{\cal X})$ be a measurable space. An {\bf interadditive partition} of $X$ is a finite measurable partition ${\cal Q}$ of $X$ such that for every $A\in {\cal X}$ 
$$ \mu (A)=\sum_{P\in {\cal Q}} \mu (P\cap A).$$  
\end{definition}

Then, the following holds.

\begin{proposition}\cite{musufu97}
Let $(X, {\cal X})$ be a measurable space and ${\cal Q}$ be a finite measurable partition of $X$. Then, ${\cal Q}$ is an interadditive partition if and only if for every measurable function $f$ 
\begin{equation}\label{intchoiec}
({\cal C})\int_X f d\mu = \sum_{P\in {\cal Q}} ({\cal C}) \int_P fd\mu .
\end{equation}
\end{proposition}

A more general hierachical Choquet integral model based in what is called {\it inclusion-exclusion coverings} appears in \cite{sufumu95}.

\section{$p$-symmetric measures}
\noindent

Let us consider an OWA operator. If we look at the definition, we
can see that only the order in the scores is important, i.e. we are interested in the scores, but we do not care about which criterium each score has been obtained. Mathematically, this means that the fuzzy measure defining
the OWA operator only depends on the cardinality of the subsets,
and not in the elements of the subset themselves.

Thus, all criteria have the same importance or, in other words, we have a ``subset of indifference'' ($X$ itself). Then, it makes sense to define 2-symmetric measures as those
measures for which we have two subsets of indifference,
3-symmetric measures as those with three subsets of indifference,
and so on. Let us now translate this idea.

\begin{definition}
Given two elements $x_i, x_j$ of the universal set $X$, we say
that $x_i$ and $x_j$ are {\bf indifferent elements} if and only if
$$ \forall A\subset X\backslash \{ x_i, x_j\} ,\, \mu (A\cup
x_i)=\mu (A\cup x_j).$$
\end{definition}

This definition translates the idea that we do not care about
which element, $x_i$ or $x_j$, is in the coalition; that is, we are indifferent
between $x_i$ and $x_j$. This concept can be generalized for subsets of
more than two elements, as shown in the following definition:

\begin{definition}\label{subindif}
Given a subset $A$ of $X$, we say that $A$ is a {\bf set of
indifference} if and only if $$ \forall B_1, B_2\subset A,\,
|B_1|=|B_2|,\, \forall C\subset X\backslash A,\,  \mu (B_1\cup
C)=\mu (B_2\cup C).$$
\end{definition}

It is easy to see the following:

\begin{lemma}\label{altdef}
Given $A\subset X$, $A$ is a set of indifference if and only if
$$ \forall B_1, B_2\subset A,\, |B_1|=|B_2|,\, \forall C\subset
X\backslash \{ B_1\cup B_2\} ,\, \mu (B_1\cup C)=\mu (B_2\cup
C).$$
\end{lemma}

{\bf Proof:} For $C\subset X\backslash A$, we have, applying Definition \ref{subindif}, $ \mu (C\cup B_1)=\mu (C\cup B_2).$

Let us consider $C\subset X\backslash (B_1\cup B_2)$ but $C\not\subset X\backslash A$. Then, $\exists D\subset A\backslash (B_1\cup B_2)$ such that $C=D\cup C'$, with $C'\subset X\backslash A.$ Thus, by Definition \ref{subindif},
$$ \mu (C\cup B_1)=\mu (C'\cup D\cup B_1)=\mu (C'\cup D\cup B_2)=\mu (C\cup B_2),$$ and therefore the result holds. $\cqd $ 
 
Another property of sets of indifference is:

\begin{lemma}\label{indif}
If $A$ is a set of indifference and $A'\subset A$, then $A'$ is itself a set of indifference.
\end{lemma}


\begin{example}
Consider $A,$ a set of indifference. Then, taking $C=\emptyset
$, we obtain $$ \mu (x_i)=\mu (x_j),\, \forall x_i, x_j\in A.$$ $$
\mu (x_i, x_j)=\mu (x_k, x_l),\, \forall x_i, x_j, x_k, x_l \in
A,$$ and so on.
\end{example}

An example of sets of indifference are null sets, defined in \cite{aush74} and \cite{musu91}:

\begin{definition}
A subset $A\subset X$ is called a {\bf null set} with respect to $\mu $ if $$ \mu (A\cup B)=\mu (B),\, \forall B\subset X\backslash A.$$ \end{definition}

A consequence of Definition \ref{subindif} is:

\begin{lemma}
A null set is a set of indifference.
\end{lemma}

{\bf Proof:} Let $A$ be a null set. Then, $ \mu (A\cup B)=\mu (B),\, \forall B\subset X.$
Let us consider now $A_1, A_2 \subset A,\, |A_1|=|A_2|.$ For $B\subset X\backslash A$ 
$$ \mu (B)\leq \mu (A_1\cup B)\leq \mu (A\cup B)=\mu (B).$$
$$ \mu (B)\leq \mu (A_2\cup B)\leq \mu (A\cup B)=\mu (B).$$
Then, $ \mu (A_1\cup B)=\mu (B)=\mu (A_2\cup B)$ and hence, $A$ is a set of indifference. $\cqd $

We are now able to define $p$-symmetric fuzzy measures
($p$-symmetric measures for short). We start with 2-symmetric measures.

\begin{definition}
Given a fuzzy measure $\mu $, we say that $\mu $ is a {\bf
2-symmetric measure} if and only if there exists a partition of
the universal set $\{ A, A^c\} , A, A^c\not= \emptyset $ such that
both $A$ and $A^c$ are sets of indifference and $X$ is not a set of indifference.
\end{definition}

For the general case we have:

\begin{definition}
Given two partitions $\{ A_1,..., A_p\} ,\, \{ B_1, ..., B_r\} $ of a referential $X,$ we say $\{ A_1,..., A_p\} $ is {\bf coarser} than $\{ B_1, ..., B_r\} $ if the following holds: $$ \forall A_i,\, \exists B_j \mbox{ such that } B_j\subset A_i.$$
\end{definition}

\begin{definition}
Given a fuzzy measure $\mu $, we say that $\mu $ is a {\bf
$p$-symmetric measure} if and only if the coarsest partition of
the universal set in sets of indifference is $\{ A_1,..., A_p\} , A_i\not= \emptyset ,\forall
i\in \{ 1,...,p\} $.
\end{definition}

Note that by Lemma \ref{indif} we need to work with the coarsest partition. Otherwise, a $p$-symmetric measure would be also a $p'$-symmetric measure for any $p'>p$.
 
For the 2-symmetric case, we will use both $\{ A_1, A_2\} $ and $\{ A, A^c\} $ for denoting the partition of $X$ in sets of indifference.

With these definitions, a symmetric measure is just a 1-symmetric
measure.

\begin{example}
Consider the 2-symmetric case. Consider the partition given by $A=\{
x_1,..., x_k\} ,\, A^c=\{ x_{k+1},...,x_n\} $, with $A, A^c$ two
sets of indifference. Then, in order to define the fuzzy
measure we just need to know $$ \mu (x_1), \mu (x_{k+1}), \mbox{
for singletons.} $$ $$\mu (x_1, x_2), \mu (x_{k+1}, x_{k+2}), \mu
(x_1, x_{k+1}), \mbox{ for sets of two elements},$$ and so on.

Then, it suffices to know the cardinality and the number of elements of $A$ in the subset.
\end{example}

\begin{remark}
It is important to note that, in order to define a $p$-symmetric measure, we need
to know which are the
sets of indifference partitioning the universal set. For
symmetric measures, we have only one set of indifference ($X$) and
thus we omit it, but a symmetric measure is a very particular measure and this does not hold for the general
$p$-symmetric case.
\end{remark}

Let us now propose a situation in which $p$-symmetric measures may appear:

\begin{example}
Suppose that a jury of four members is evaluating some students. Moreover, suppose that two members of the jury are mathematicians $M_1, M_2$  and the other two are physicists $P_1, P_2$. Suppose also that we do not have information about which one of the two mathematicians is the best, nor for the physicists. However, let us suppose that, for us, the marks in Mathematics are more important than those in Physics. The fuzzy measure could be defined as follows: $ \mu (M_i)=0.3,\, \mu (P_i)=0.2,\, i=1,2 $ as the marks in Mathematics are more important than the marks in Physics. Now, for pairs, we can define $ \mu (M_1, M_2)=0.5, \mu (P_1, P_2)=0.3, \mu (M_i, P_j)=0.8$. This is due to the fact that a student should be considered better (in the sense of more complete) if he obtains a good evaluation for both subjects than in the case in which he is very good in just one of them. Finally, we can define $\mu (M_1, M_2, P_i)=0.9, \mu (P_1, P_2, M_i)=0.85,\mu (X)=1$. 

In this example, we have two sets of indifference, one for the mathematicians and another one for the physicists, and $\mu $ is a 2-symmetric measure. These subsets model the fact that we are not able to distinguish between the mathematicians nor between the physicists. Then, for example, the coalition between a physicist and a mathematician has always the same importance for us, regardless which is the mathematician and the physicist in it.   
\end{example}

\begin{example}
Consider a finite referential set $X$ on which a probability measure has been defined. However, suppose that we only know the probability values on some subsets of $X$, namely $B_1, ..., B_p.$ Then, we have a set of coherent probabilities with this information. The lower bound of this set is given by $$ \mu (A)=\sum_{B_i\subset A} P(B_i),$$ and similarly, the upper bound is the corresponding dual measure. This concepts has been introduced by De Finetti in \cite{fin74}. 

Let us suppose now that sets $B_1, ..., B_p$ determines a partition on $X$. Then, it is easy to see that the corresponding measure $\mu $ is at most a $p$-symmetric measure, where sets of indifference are $B_1, ..., B_p$. Indeed, the $p$-symmetric measure is given by 
$$ \mu (i_1, ..., i_p)=\left\{ \begin{array}{ll} 0 & \mbox{ if } i_k<|B_k|,\, \forall k \\
 P(B_{i_r}) &   \mbox{ if } i_r=|B_r|,\, i_k<|B_k|,\, \forall k\not= r \\
... & ...\end{array} \right. $$
\end{example}

\begin{example}
Consider the 2-step Choquet integral defined in \cite{mevi99}. Proposition 2 speaks about 2-step
Choquet integral with second step based on additive measures which can be
represented as a single Choquet integral with respect to a fuzzy measure; now, if the first
steps are OWA operators, i.e. Choquet integral with respect to 1-symmetric fuzzy measures
$\mu_1 ,..,\mu_p$ with disjoint supports, we have that the corresponding $\mu $ in Choguet
integral representation is a $p$-symmetric measure.
\end{example}

In the following, as we only need to know the number of elements
of each set of indifference that belong to a given subset $C$
of the universal set $X$, when dealing with a $p$-symmetric measure defined by the partition $\{ A_1, ..., A_p\} $, we use the notation $C\equiv (c_1,...,
c_p),$ where $c_i$ is the number of elements of $A_i$ in $C$. Then, we can identify the different
subsets with $p$-dimensional vectors whose $i$-th coordinate is an integer
number from 0 to $|A_i|$. Hence, the number of different subsets $C$ is $(|A_1|+1)\times ... \times (|A_p|+1),$ and this is the number of necessary values that we need to know to completely determine the $p$-symmetric fuzzy measure. Moreover, as $\mu (0,..., 0)=0, \mu (|A_1|, ...., |A_p|)=1$, it follows that we only need to determine $(|A_1|+1)\times ... \times (|A_p|+1)-2$ values. This is written in next proposition:

\begin{proposition}\label{numsub}
Let $\mu $ be a $p$-symmetric measure with respect to the partition $ \{ A_1,..., A_p\} $. Then, the number of values that are needed in order to determine $\mu $ is
$$ \big[ (|A_1|+1)\times \cdots \times (|A_p|+1)\big]  -2 .$$ 
\end{proposition}

\begin{example}\label{ex2syA1}
Consider the special 2-symmetric case in which $A=\{ x_1\} $.
Then, in order to define the fuzzy measure, we just need to know $$
\mu (x_1), \mu (x_2), \mu (x_1, x_2), \mu (x_2, x_3), ..., \mu
(x_1,..., x_{n-1}), \mu (x_2,...,x_n),$$ or, in the notation proposed before $$ \mu (1,0), \mu (0,1), ..., \mu (1,n-2), \mu (0, n-1),$$ i.e. $2n-2$ values.
\end{example}

\begin{remark}
Note that the number of different subsets depends not only on the
degree of symmetry, but also on the sets of indifference that
determine the partition of $X$. In Example \ref{ex2syA1}, we only
needed $2n-2$ coefficients. However, if we take $n=6, |A_1|=3, |A_2|=3$, by Proposition \ref{numsub}, we need 4*4-2=14 coefficients.
\end{remark}

As a consequence of Proposition \ref{numsub}, a $p$-symmetric fuzzy measure can
be represented in a $(|A_1|+1)\times ... \times (|A_p|+1)$ matrix
${\bf M}$ such that ${\bf M} [c_1,..., c_p]=\mu (c_1,..., c_p)$.

Let us see some special cases as examples:

\begin{itemize}
\item
If we have a 1-symmetric measure, we just need to know a
$(n+1)$-dimensional vector $\vec{v}$ such that $\vec{v}(i)=\mu
(i),$ where $\mu (0)=0$ and $\mu (n)=1$.
\item
If we have a 2-symmetric measure, we obtain a $(|A|+1)\times
(|A^c|+1)$ matrix.
{\small $$ \left( \begin{array}{ccccc} \mu (0,0) & \mu (0,1) & \hdots  & \mu (0,|A_2|-1) & \mu (0,|A_2|) \\
       \mu (1,0) & \mu (1,1) & \hdots  & \mu (1,|A_2|-1) & \mu (1,|A_2|) \\
       \vdots    & \vdots    & \ddots  & \vdots          & \vdots        \\
       \mu (|A_1|-1,0) & \mu (|A_1|-1,1) & \hdots  & \mu (|A_1|-1,|A_2|-1) & \mu (|A_1|-1,|A_2|) \\
       \mu (|A_1|,0) & \mu (|A_1|,1) & \hdots  & \mu (|A_1|-1,|A_2|-1) & \mu (|A_1|,|A_2|) \end{array} \right) $$}
\item
In the extreme case of a $n$-symmetric measure, we obtain a
$2\times ...\times 2$ matrix, i.e. we need $2^n$ coefficients (two of them are $\mu (\emptyset )$ and $\mu (X)$).
\end{itemize}

We finish this section with the following result related to dual measures.

\begin{lemma}
Let $\mu $ be a $p$-symmetric measure with respect to a partition $\{ A_1, ..., A_p\} $. Then, $\bar{\mu }$ is also a $p$-symmetric measure with respect to the same partition. 
\end{lemma}

{\bf Proof:} Let us consider a $p$-symmetric measure $\mu $ with respect to the partition $\{ A_1,..., A_p\} .$ To show that $\bar{\mu }$ is another $p$-symmetric measure, it suffices to note that $$ \bar{\mu } (i_1,..., i_p)=1-\mu (|A_1|-i_1, ...,|A_p|-i_p), $$
whence the result holds. \cqd 
 
\section{Other representations of a $p$-symmetric measure}
\noindent

In this section, we deal with the problem of obtaining the
different representations of a fuzzy measure in the special case
of $p$-symmetric measures. More concretely, we obtain the M\"obius
transform and the Shapley interaction.

Let us start with the M\"obius transform.

\begin{proposition}\label{mobmeapsym}
Let $\mu $ be a $p$-symmetric measure associated to the partition
$\{ A_1,...,A_p\} $. Then, for $B\equiv (b_1,..., b_p)\subset X$,
we have $$ m(b_1,...,b_p)=\sum_{i_1\leq b_1,...,i_p\leq b_p}
(-1)^{b_1+...+b_p-i_1-...-i_p}{b_1\choose i_1}...{b_p\choose i_p}
\mu (i_1,...,i_p).$$
\end{proposition}

{\bf Proof:} Consider $B\equiv (b_1,..., b_p)$. Then, the number of subsets of
$B$ with $i_1$ elements of $A_1$, $i_2$ elements of $A_2$..., $i_p$ elements of $A_p$ is
$$  {b_1\choose i_1}...{b_p\choose i_p}, $$ and we know that they
have all the same measure.

Now, as
$$ m(B)=\sum_{C\subset B}(-1)^{|B|-|C|}\mu (C),$$
by (\ref{defmob}), we obtain
$$ m(b_1,...,b_p)=\sum_{i_1\leq b_1,...,i_p\leq b_p} (-1)^{b_1+...+b_p-i_1-...-i_p}{b_1\choose i_1}...{b_p\choose i_p} \mu (i_1,...,i_p), $$ whence the result. $\cqd $

Let us now find the expression of the measure in terms of the M\"obius
transformation.

\begin{proposition}\label{meamobpsym}
Let $\mu $ be a $p$-symmetric measure associated to the partition
$\{ A_1,..., A_p\} $. Now, suppose $m$ denotes its M\"obius
transform. Then, for $B\equiv (b_1,..., b_p)\subset X$, it is $$
\mu (b_1,..., b_p)=\sum_{c_1\leq b_1,...,c_p\leq b_p} {b_1\choose
c_1}\cdots {b_p\choose c_p} m(c_1,..., c_p).$$
\end{proposition}

{\bf Proof:} Consider $C\equiv (c_1,..., c_p)\subset B$. Then, the number of
possibilities for such a $C$ is $$ {b_1\choose c_1}\cdots {b_p\choose c_p},$$ and
thus the expression holds applying (\ref{mob2mea}). $\cqd $

Let us now turn to the Shapley interaction:

\begin{proposition}\label{shamobpsym}
Let $\mu $ be a $p$-symmetric measure associated to the partition
$\{ A_1,..., A_p\} $. Then, for $B\equiv (b_1,..., b_p)\subset X,$
we have $$ I(b_1,..., b_p)=\sum_{c_1\geq b_1,..., c_p\geq
b_p}{1\over c-b+1}{a_1-b_1\choose c_1-b_1}\cdots
{a_p-b_p\choose c_p-b_p}m(c_1,..., c_p),$$ with $c=\sum_{i=1}^p c_i,\, b=\sum_{i=1}^p b_i.$
\end{proposition}

{\bf Proof:} We know from (\ref{mob2sha}) that for $B\subset X$ $$ I(B)=\sum_{C|B\subset C} {1\over
|C|-|B|+1}m(C).$$ Let us consider $C\equiv (c_1,...,c_p)|\, B\equiv
(b_1,...,b_p)\subset C$. Then, the number of possible $C$'s is $$
{a_1-b_1\choose c_1-b_1}\cdots {a_p-b_p\choose c_p-b_p}.$$ Thus,
we obtain $$ I(b_1,..., b_p)=\sum_{c_1\geq b_1,..., c_p\geq
b_p}{1\over c-b+1}{a_1-b_1\choose c_1-b_1}\cdots
{a_p-b_p\choose c_p-b_p}m(c_1,..., c_p), $$ whence the result. $\cqd $

The reciprocal result is given by 

\begin{proposition}\label{shap2mobsym}
Let $\mu $ be a $p$-symmetric measure associated to the partition
$\{ A_1,..., A_p\} $. Then, for $B\equiv (b_1,..., b_p)\subset X,$
we have $$ m(b_1,..., b_p)=\sum_{c_i\leq a_i-b_i, i=1,..., p}{a_1-b_1\choose c_1}\cdots
{a_p-b_p\choose c_p} B_{c_1+...+c_p}I(c_1+b_1,..., c_p+b_p).$$
\end{proposition}

{\bf Proof:} We know from (\ref{sha2mob}) that for $B\subset X$ $$ m(B)=\sum_{C\subset X\backslash B} B_{|C|}I(C\cup B).$$ Let us consider $C\equiv (c_1,...,c_p)\subset X\backslash B\equiv
(a_1-b_1,...,a_p-b_p)$. Then, the number of possible $C$'s is $$
{a_1-b_1\choose c_1}\cdots {a_p-b_p\choose c_p}.$$ Thus,
we obtain $$ m(B)=\sum_{c_1\leq a_1-b_1,..., c_p\leq a_p-
b_p}B_{c_1+...+c_p}{a_1-b_1\choose c_1}\cdots
{a_p-b_p\choose c_p} I(c_1+b_1,..., c_p+b_p), $$ whence the result. $\cqd $

The expression of Shapley interaction in terms of $\mu $ is given
in next proposition.

\begin{proposition}
Let $\mu $ be a $p$-symmetric measure associated to the partition
$\{ A_1,..., A_p\} $. Then, for $B\equiv (b_1,..., b_p)\subset X,$
we have {\small $$ I(B)=\sum_{d_i\leq a_i, \forall i}\, 
\sum_{c_i\geq \{ b_i, d_i\} ,\forall i} {1\over
c-b+1}(-1)^{c-d} {c_1\choose d_1}\cdots {c_p\choose
d_p}{a_1-b_1\choose c_1-b_1}\cdots {a_p-b_p\choose c_p-b_p}\mu
(D),$$} with $\displaystyle{d=\sum_{i=1}^p d_i, c=\sum_{i=1}^p c_i, b=\sum_{i=1}^p b_i}.$ 
\end{proposition}

{\bf Proof:} We will use the expressions in Proposition \ref{mobmeapsym} and Proposition \ref{shamobpsym}. From Proposition \ref{shamobpsym}, we know that $$  I(b_1,..., b_p)=\sum_{c_1\geq b_1,..., c_p\geq
b_p}{1\over c-b+1}{a_1-b_1\choose c_1-b_1}\cdots
{a_p-b_p\choose c_p-b_p}m(c_1,..., c_p). $$ But now, from Proposition \ref{mobmeapsym}, $$ m(c_1, ..., c_p)=\sum_{d_1\leq c_1,...,d_p\leq c_p}
(-1)^{c_1+...+c_p-d_1-...-d_p}{c_1\choose d_1}...{c_p\choose d_p}
\mu (d_1,...,d_p).$$ Joining both results, the proposition is proved. $\cqd $

And the reciprocal result is:

\begin{proposition}\label{mobshapsym}
Let $\mu $ be a $p$-symmetric measure associated to the partition
$\{ A_1,..., A_p\} $. Then, for $B\equiv (b_1,..., b_p)\subset X$
we have {\small $$ \mu (B)=\sum_{c_i\leq b_i, \forall i}{b_1\choose c_1} \dots {b_p\choose c_p}
\sum_{d_i\leq a_i- c_i ,\forall i} {a_1-c_1\choose d_1}\cdots {a_p-c_p\choose
d_p}B_{d}I (c_1+d_1,..., c_p+d_p),$$} with $d=\sum_{i=1}^p d_i.$
\end{proposition}

{\bf Proof:} We know that
$$ \mu (b_1,..., b_p)=\sum_{c_1\leq b_1,...,c_p\leq b_p} {b_1\choose
c_1}\cdots {b_p\choose c_p} m(c_1,..., c_p),$$ by Proposition \ref{meamobpsym}, and 
$$  m(C)=\sum_{d_i\leq a_i-c_i, i=1,..., p}{a_1-c_1\choose d_1}\cdots
{a_p-c_p\choose d_p} B_{d_1+...+d_p}I(c_1+d_1,..., c_p+d_p),$$ by Proposition \ref{shap2mobsym}. Joining both results, the proposition is proved. $\cqd $

Of course, when considering the representation of a $p$-symmetric measure in terms of the M\"obius transform or the Shapley interaction, we can represent it in a $p$-dimensional matrix, as we have done in the previous section.

\section{Choquet integral with respect to a $p$-symmetric measure}
\noindent

In this section we study the expression of Choquet integral with
respect to a $p$-symmetric measure, as well as some properties of
this integral.

\begin{proposition}
Let $\mu $ be a $p$-symmetric measure. Given a function $f$, the Choquet integral is given by $$ \sum_{i=1}^n f(x_{(i)})\sum_{c_k\leq b_k^{i-1},\forall k}  m(c_1,..., c_j+1,..., c_p)\prod_{k=1}^p {b_k^{i-1}\choose c_k}$$ with $ x_{(i)}\in A_j,$ and where $(b_1^{i-1},...,b_p^{i-1})\equiv B_{(i-1)}=\{ x_{(1)},..., x_{(i-1)}\} .$
\end{proposition}

{\bf Proof:} We know from Definition \ref{choint}
$$ (C)\int fd\mu =\sum_{i=1}^n f(x_{(i)})(\mu (B_{(i)})-\mu (B_{(i+1)})),$$
where $B_{(i)}=B_{(i+1)}\cup \{ x_{(i)}\} .$

Suppose that $B_{(i+1)}\equiv (b_1^{i-1},..., b_p^{i-1})$. Then,
by Proposition \ref{meamobpsym} $$ \mu (B_{(i+1)})=\sum_{c_k\leq
b_k^{i-1},\forall k} m(c_1,..., c_p) \prod_{k=1}^p {b_k^{i-1}\choose c_k}.$$ Now, if $x_{(i)}\in A_j$, {\small $$
\mu (B_{(i)})-\mu (B_{(i+1)})=\sum_{C\subset B_{(i+1)}} m(x_{(i)}\cup
C)=\sum_{c_k\leq b_k^{i-1},\forall k}  m(c_1,...,
c_j+1,..., c_p)\prod_{k=1}^p {b_k^{i-1}\choose c_k},$$} whence the result. $\cqd $

\begin{figure}[htb]
\begin{center}
$
\epsfxsize =6cm
 \epsfbox{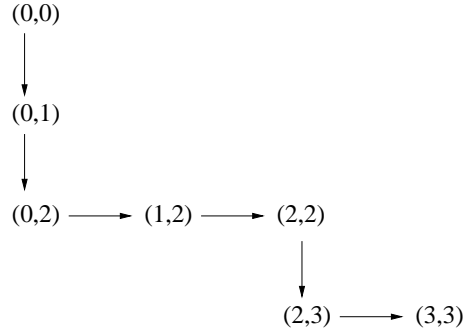}
 $
\end{center}
\caption{Possible path from (0,0) to (3,3) when $|A_1|=3$ and $|A_2|=3$.}
\label{fig2:path}
\end{figure}

As a subset $C\subset X$ is determined by the number of elements in $A_i,\forall i$, we can find all possible
Choquet integrals finding all possible paths from $(0,..., 0)$ to
$(|A_1|,..., |A_p|)$ (see Figure \ref{fig2:path} for an example with a 2-symmetric measure).

The number of such paths is given in next lemma:

\begin{lemma}
Let $\mu $ be a $p$-symmetric measure with respect to the partition $\{ A_1,..., A_p\} $. Then, the number of paths from $(0,..., 0)$ to $(|A_1|,..., |A_p|)$ is 
$$ {n\choose |A_1|,..., |A_p|}.$$
\end{lemma}


\begin{example}
If we are in the 2-symmetric case and $|A|=1$, then we have just $n+1$ different paths from $(0,...,
0)$ to $(|A_1|,..., |A_p|)$ (see Figure \ref{fig:path}).
\begin{figure}[htb]
\begin{center}
$
\epsfxsize =4cm
 \epsfbox{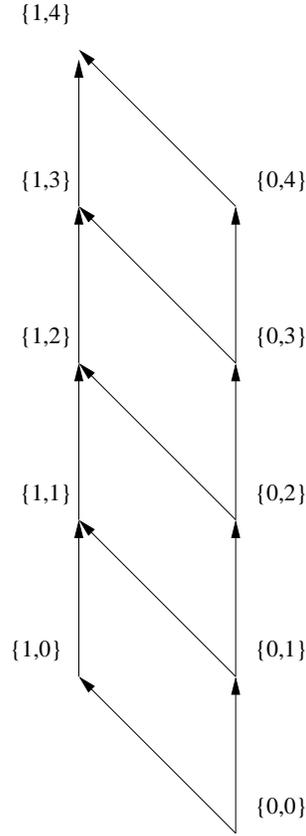}
 $
\end{center}
\caption{Possible paths when $|A_1|=1$ and $|A_2|=4$.}
\label{fig:path}
\end{figure}
\end{example}

Let us now see some properties for Choquet integral of a
$p$-symmetric measure.

\begin{proposition}\label{choowapsym}
Let $\mu $ be a $p$-symmetric measure with respect to the
partition $\{ A_1,..., A_p\} ,$ and suppose $\mu (A_i)>0,\forall i$. Then, the Choquet integral is given by $$ \sum_{i=1}^p\mu (A_i)({\cal C})\int f d\mu_{A_i} +\sum_{B\not\subset
A_j,\forall j}m(B)\bigwedge_{x_i\in B} f(x_i),$$ where $\mu_{A_i}$ is defined by its M\"obius transform \[ m_{A_i}(C)=\left\{\begin{array}{ll} {m (C)\over \mu (A_i)} & {\rm if }
C\subset A_i \\ 0 & {\rm otherwise} \end{array} \right. \] 
\end{proposition}

{\bf Proof:} Suppose that $\mu $ is a $p$-symmetric measure with respect to the
partition $\{ A_1,..., A_p\} $. Then, the Choquet integral can be written as $$ \sum_{j=1}^p \sum_{B\subset A_j}m(B)\bigwedge_{x_i\in B} f(x_i)+
\sum_{B\not\subset A_j,\forall j}m(B)\bigwedge_{x_i\in B}
f(x_i),$$ by (\ref{chomob}). Now, let us define for $A\subset X,\, \mu (A)>0$  \[  m_A(C)=\left\{ \begin{array}{ll} {m (C)\over \mu (A)} & {\rm if }
C\subset A \\ 0 & {\rm otherwise} \end{array}\right. \]  Let us see that $m_A$ is
the M\"obius transform of a fuzzy measure. To see this, let us show that the conditions of Proposition \ref{monconmob} hold:

First, note that $m_A(\emptyset )=0$. Now, for $i\in X, C\subset
X$, we have

\begin{itemize}
\item
If $x_i\notin A$, then
$$ \sum_{x_i\in B\subset C} m_A(B)=0.$$
\item
If $x_i\in A$, then
$$ \sum_{x_i\in B\subset C} m_A(B)= \sum_{x_i\in B\subset C\cap A} {m(B)\over \mu (A)}\geq 0,$$
as $\mu $ is a fuzzy measure.
\item
$ \sum_{B\subset X} m_A(B)={\mu (A)\over \mu (A)}=1.$
\end{itemize}

Let us denote by $\mu_A $ the fuzzy measure associated to $m_A$.
Then, it is trivial to see that $$\sum_{B\subset
A_j}m(B)\bigwedge_{x_i\in B} f(x_i)=\mu (A_j)\, (C)\int fd\mu_{A_j},\, \forall
j,$$ 
by (\ref{chomob}). This completes the proof. $\cqd $

The last summand in Proposition \ref{choowapsym} represents the
part of the Choquet integral that cannot be assigned to any
subset in the partition. When $\mu $ is a belief function, the
following can be proved:

\begin{proposition}
Let $\mu $ be a $p$-symmetric measure with respect to the
partition $\{ A_1,..., A_p\} $. Suppose also that $\mu $ is a
belief function. Then, the Choquet integral can be written as $$ \sum_{i=1}^n\mu (A_i)({\cal C})\int f d\mu_{A_i}+(C)\int
fd\mu^*,$$ where $\mu_{A_i}$ and $\mu^*$ are defined by
\[ m_{A_i}(C)=\left\{ \begin{array}{ll} {m (C)\over \mu (A_i)} & {\rm if } C\subset A_i \\ 0 & {\rm otherwise} \end{array} \right. \] 
$$ \mu^* (C)=\mu (C)-\mu (C\cap A_1)-...- \mu
(C\cap A_p).$$
\end{proposition}

{\bf Proof:} We know from Proposition \ref{choowapsym} that the Choquet integral for a $p$-symmetric fuzzy measure can be written as $$ \sum_{i}\mu (A_i)({\cal C})\int f d\mu_{A_i} +\sum_{B\not\subset
A_j,\forall j}m(B)\bigwedge_{x_i\in B} f(x_i).$$ Now, define $$
\mu^* (C)=\mu (C)-\mu (C\cap A_1)-...- \mu (C\cap A_p).$$ $\mu^* $
is a non-normalized fuzzy measure:
$$ \sum_{x_i\in B\subset C} m^*(B)=\sum_{x_i\in B\subset C,\, B\not\subset A_j,\forall j} m(B)\geq 0,$$ as $\mu $ is a belief function. Remark that
\[ m^*(B)=\left\{ \begin{array}{ll} m(B) & {\rm if } B\not\subset A_j,\forall j \\ 0 & {\rm otherwise} \end{array}\right. \] 
Then, it is easy to see that $$ \sum_{B\not\subset A_j,\forall
j}m(B)\bigwedge_{x_i\in B} f(x_i)=(C)\int fd\mu^*,$$ and thus, the
proposition holds. $ \cqd $

Note that for belief functions, $\mu^* (X)=\mu
(X)-\mu (A_1)-...- \mu (A_p).$ Then, this value can be seen
as a degree of the interaction among the elements of the
partition: If $\mu^*(X)=0$, then necessary $m(B)=0$ if $B\not\subset A_1, ..., B\not\subset A_p.$ We write it in the following definition.

\begin{definition}
Consider $\mu $ a $p$-symmetric measure associated to the
partition given by $\{ A_1,..., A_p\} $. Suppose also that $\mu $ is a
belief function. We define the {\bf degree of interaction among the
elements of the partition} by $$\mu (X)-\mu (A_1)-...-
\mu (A_p).$$
\end{definition}

Now, we can state the following corollary:

\begin{corollary}\label{lastcor}
Let $\mu $ be a $p$-symmetric measure with respect to the
partition given by $\{ A_1,..., A_p\} $ such that $\mu (A_i)>0$. Suppose also that $\mu $ is a
belief function. When the interaction degree vanishes, the Choquet integral can be written as $$ \mu (A_1)({\cal C})\int f d\mu_{A_1}+...+\mu (A_p)({\cal C})\int f d\mu_{A_p},$$
where $\mu_{A_i}$ is defined by \[ m_{A_i}(C)=\left\{ \begin{array}{ll}{m
(C)\over \mu (A_i)} & {\rm if } C\subset A_i \\ 0 & {\rm otherwise} \end{array}\right. \]  
\end{corollary}

In this sense, when $\mu $ is a belief function and the degree of interaction
vanishes, the partition $\{ A_1,..., A_p\} $ in sets of indifference is also an interadditive partition (Equation \ref{intchoiec}). Moreover, each integral is indeed an OWA operator over a smaller referential set.

\begin{remark}
Note that we need $\mu $ being a belief function in order to
ensure a positive value for the degree of interaction among the
elements of the partition. Moreover, if $\mu $ is not a belief function, we can find $\mu (X)-\mu (A_1)- ...- \mu (A_p)=0$ and, on the other hand, there exist interactions among the elements of the partition. 
\end{remark}

For Corollary \ref{lastcor}, it must be remarked that $\mu $ must be a belief function. Otherwise, the result does not necessary hold as next example shows:

\begin{example}
Consider $X=\{ x_1, x_2, x_3\} $ and let us define the fuzzy measure $\mu $ given by the following M\"obius transform:

\begin{center}
\begin{tabular}{|c|c|c|c|c|c|c|} 
\hline $ x_1$ & 
$x_2$ & $ x_3$ & $ x_1, x_2$ & $ x_1,x_3$ & $ x_2, x_3$ & $x_1, x_2, x_3 $ \\
\hline 0.4 & 0.3 & 0.3 & 0.1 & 0.1 & 0 & -0.2 \\ \hline 
\end{tabular}
\end{center}

$\mu $ is a 2-symmetric measure, with sets of indifference $A_1=\{ x_1\} ,\, A_2=\{ x_2, x_3\} .$ On the other hand $\mu (A_1)+\mu (A_2)=1$, and thus, $\mu (X)- \mu (A_1)-\mu (A_2)=0$. However, we can not ensure   
 $$ ({\cal C})\int f d\mu = \mu (A_1)\, ({\cal C})\int f d\mu_{A_1}+\mu (A_2)\, ({\cal C})\int f d\mu_{A_2},\, \forall f.$$ To see this, just consider $f$ defined by $f(x_1)=1,\, f(x_2)=0.5,\, f(x_3)=0.$ Then, it is straightforward to see  $$ ({\cal C})\int f d\mu = 0.6, \,  \mu (A_1)\, ({\cal C})\int f d\mu_{A_1}= 0.4,\, \mu (A_2)\, ({\cal C})\int f d\mu_{A_2}=0.15.$$
\end{example}

\section{Conclusions}
\noindent

In this paper, we have proposed a generalization of the concept of symmetry for fuzzy measures. This new concept is based in sets of indifference; these subsets model the fact that some elements are indistinguishable. We have defined $p$-symmetric fuzzy measures and we have studied some of their properties, as well as other representations. The main property of $p$-symmetric measures is that they can be represented in a $p$-dimensional matrix. Once the definition of $p$-symmetry given, we have obtained an expression for Choquet integral; we have shown that this integral can be easily computed from the matrix representation. Finally, we have derived a value for the interaction among sets of indifference. 

We think that $p$-symmetric measures provide an interesting tool in the field of fuzzy measures, and a graduation between symmetric measures and fuzzy measures.

%

\section{Acknowledgements}

The research in this paper has been supported in part by FEDER-MCYT grant number BFM2001-3515.
 
{\small 
\bibliographystyle{plain}

\bibliography{/home/pedro/Biblio/ae.bib,/home/pedro/Biblio/fj.bib,/home/pedro/Biblio/kn.bib,/home/pedro/Biblio/ot.bib,/home/pedro/Biblio/uz.bib}

\begin{thebibliography}{10}

\bibitem{aush74}
R.~J. Aumann and L.~S. Shapley.
\newblock {\em Values of Non-Atomic Games}.
\newblock Princeton University Press, 1974.

\bibitem{capr01}
T.~Calvo and A.~Pradera.
\newblock Some characterizations based on double aggregation operators.
\newblock In {\em Proceedings of European Society for Fuzzy Logic and
  Technology (EUSFLAT) Meeting}, pages 470--474, Leicester (UK), September
  2001.

\bibitem{chco00}
A.~Chateauneuf and M.~Cohen.
\newblock Choquet {E}xpected {U}tility {M}odel: {A} new approach to individual
  behaviour under uncertainty and to {S}ocial {W}elfare.
\newblock In M.~Grabisch, T.~Murofushi, and M.~Sugeno, editors, {\em Fuzzy
  {M}easures and {I}ntegrals}, pages 289--314. Physica-{V}erlag, 2000.

\bibitem{chja89}
A.~Chateauneuf and J.-Y. Jaffray.
\newblock Some characterizations of lower probabilities and other monotone
  capacities through the use of {M}\"obius inversion.
\newblock {\em Mathematical Social Sciences}, (17):263--283, 1989.

\bibitem{cho53}
G.~Choquet.
\newblock Theory of capacities.
\newblock {\em Annales de l'Institut Fourier}, (5):131--295, 1953.

\bibitem{dem67}
A.~P. Dempster.
\newblock Upper and lower probabilities induced by a multivalued mapping.
\newblock {\em Ann. Math. Statist.}, (38):325--339, 1967.

\bibitem{dehape97}
J.~Derks, H.~Haller, and H.~Peters.
\newblock The selectope for cooperative games.
\newblock Technical Report RM/97/016, Univ. Maastricht, 1997.

\bibitem{fin74}
D.~De Finetti.
\newblock {\em Theory of {P}robability. {V}olume 1}.
\newblock Wiley Classics Library, 1974.

\bibitem{gra96b}
M.~Grabisch.
\newblock Fuzzy measures and integrals: A survey of applications and recent
  issues.
\newblock In D.~Dubois, H.~Prade, and R.~Yager, editors, {\em Fuzzy sets
  methods in information engeneering: A guide tour of applications}. 1996.

\bibitem{gra96c}
M.~Grabisch.
\newblock $k$-order additive discrete fuzzy measures.
\newblock In {\em Proceedings of 6th Int. Conf. on Information Processing and
  Management of Uncertainty in Knowledge-Based Systems (IPMU)}, pages
  1345--1350, Granada (Spain), 1996.

\bibitem{gra97c}
M.~Grabisch.
\newblock $k$-order additive discrete fuzzy measures and their representation.
\newblock {\em Fuzzy Sets and Systems}, (92):167--189, 1997.

\bibitem{grro00}
M.~Grabisch and M.~Roubens.
\newblock Application of the {C}hoquet integral in {M}ulticriteria {D}ecision
  {M}aking.
\newblock In M.~Grabisch, T.~Murofushi, and M.~Sugeno, editors, {\em Fuzzy
  Measures and Integrals}, pages 348--375. Physica-{V}erlag, 2000.

\bibitem{halipo52}
G.~H. Hardy, J.~E. Littlewood, and G.~P\'olya.
\newblock {\em Inequalities}.
\newblock Cambridge University Press, Cambridge (UK), 1952.

\bibitem{mevi99}
R.~Mesiar and D.~Vivona.
\newblock Two-step integral with respect to fuzzy measures.
\newblock {\em Tatra Mountains Mathematical Publications}, 16:359--368, 1999.

\bibitem{migr00}
P.~Miranda and M.~Grabisch.
\newblock Characterizing $k$-additive fuzzy measures.
\newblock In {\em Proceedings of Eighth International Conference of Information
  Processing and Management of Uncertainty in Knowledge-based Systems (IPMU)},
  pages 1063--1070, Madrid (Spain), July 2000.

\bibitem{muso93}
T.~Murofushi and S.~Soneda.
\newblock Techniques for reading fuzzy measures (iii): Interaction index.
\newblock In {\em Proceedings of 9th Fuzzy System Symposium}, pages 693--696,
  Sapporo (Japan), May 1993.

\bibitem{musu91}
T.~Murofushi and M.~Sugeno.
\newblock Fuzzy t-conorm integrals with respect to fuzzy measures:
  generalization of {S}ugeno integral and {C}hoquet integral.
\newblock {\em Fuzzy Sets and Systems}, (42):57--71, 1991.

\bibitem{musu93}
T.~Murofushi and M.~Sugeno.
\newblock Some quantities represented by the {C}hoquet integral.
\newblock {\em Fuzzy Sets and Systems}, (56):229--235, 1993.

\bibitem{musufu97}
T.~Murofushi, M.~Sugeno, and K.~Fujimoto.
\newblock Separated hierarchical decomposition of {C}hoquet integral.
\newblock {\em Int. J. Uncertainty, Fuzziness and Knowledge-Based Systems},
  5:563--585, 1997.

\bibitem{rot64}
G.~C. Rota.
\newblock On the foundations of combinatorial theory {I}. {T}heory of
  {M}\"obius functions.
\newblock {\em Zeitschrift f\"ur {W}ahrscheinlichkeitstheorie und {V}erwandte
  {G}ebiete}, (2):340--368, 1964.

\bibitem{sch86}
D.~Schmeidler.
\newblock Integral representation without additivity.
\newblock {\em Proc. of the Amer. Math. Soc.}, (97(2)):255--261, 1986.

\bibitem{sha76}
G.~Shafer.
\newblock {\em A Mathematical Theory of Evidence}.
\newblock Princeton University Press, Princeton, New Jersey, USA, 1976.

\bibitem{sha53}
L.~S. Shapley.
\newblock A value for $n$-person games.
\newblock In H.~W. Kuhn and A.~W. Tucker, editors, {\em Contributions to the
  theory of Games}, volume~II of {\em Annals of Mathematics Studies}, pages
  307--317. Princeton University Press, 1953.

\bibitem{sug74}
M.~Sugeno.
\newblock {\em Theory of fuzzy integrals and its applications}.
\newblock PhD thesis, Tokyo Institute of Technology, 1974.

\bibitem{sufumu95}
M.~Sugeno, K.~Fujimoto, and T.~Murofushi.
\newblock A hierarchical decomposition of {C}hoquet integral model.
\newblock {\em International Journal of Uncertainty, Fuzziness and
  Knowledge-Based Systems}, 3(1):1--15, 1995.

\bibitem{yag88}
R.~R. Yager.
\newblock On ordered weighted averaging aggregation operators in multicriteria
  decision making.
\newblock {\em IEEE Trans. Systems, Man \& Cybern.}, (18):183--190, 1988.

\end{thebibliography}
 
}
\end{document}